\documentclass[12pt,preprint]{aastex}
\newcommand{\hd}{HD\,5980}
\newcommand{\xmm}{{\sc XMM}\emph{-Newton}}

\shorttitle{First extragalactic colliding wind X-rays}
\shortauthors{Naz\'e et al.}

\begin{document}

\title{First detection of phase-dependent colliding wind X-ray emission outside the Milky Way\footnote{Based on observations collected with \xmm, an ESA Science Mission with instruments and contributions directly funded by ESA Member States and the USA (NASA).}}

\author{Ya\"el Naz\'e\altaffilmark{2}}
\affil{Institut d'Astrophysique et de G\'eophysique,
Universit\'e de Li\`ege, All\'ee du 6 Ao\^ut 17, Bat. B5c, B4000 -
Li\`ege, Belgium}
\email{naze@astro.ulg.ac.be} 

\author{Michael F. Corcoran}
\affil{Universities Space Research Association, 10211 Wincopin Circle, Suite 500
Columbia, MD 21044-3432 and Center for Space Science and Technology, Code 662, Goddard Space Flight Center, Greenbelt, MD 20771, USA }
\email{corcoran@milkyway.gsfc.nasa.gov} 

\author{Gloria Koenigsberger}
\affil{Instituto de Ciencias Fisicas, Universidad Nacional Aut\'onoma de M\'exico, Apartado Postal 48-3, Cuernavaca, Morelos 62251, Mexico}
\email{gloria@astroscu.unam.mx} 

\and

\author{Anthony F.J. Moffat}
\affil{D\'epartement de Physique, Universit\'e de Montr\'eal, CP 6128, Succursale Centre-Ville, Montr\'eal, QC H3C 3J7, and Observatoire du Mont M\'egantic, Canada}
\email{moffat@ASTRO.UMontreal.CA}

\altaffiltext{2}{Postdoctoral Researcher F.N.R.S.}

\begin{abstract}
After having reported the detection of X-rays emitted by the peculiar system \hd, we assess here the origin of this high-energy emission from additional X-ray observations obtained with \xmm. This research provides the first detection of apparently periodic X-ray emission from hot gas produced by the collision of winds in an evolved massive binary outside the Milky Way. It also provides the first X-ray monitoring of a Luminous Blue Variable only years after its eruption and shows that the dominant source of the X-rays is not associated with the ejecta. 
\end{abstract}


\section{Introduction}
The most massive stars, which are also the most luminous (and for most of their lives the hottest) non-degenerate stellar objects, have a large impact on their host galaxies. For example, throughout their evolution, these stars convert simple into more complex elements, and distribute them into space by a complex process of mass loss. They are thus largely responsible for the chemical enrichment of the Universe \citep{mas03}. Their intense, hot photon emission is also able to ionize the surrounding medium (leading to the formation of H{\sc ii} regions) and to drive a continuous ejection of matter in the form of a stellar wind. Finally, massive stars inject a large amount of mechanical energy into their host galaxies, through e.g. supersonic winds, large transient eruptions (like those of Luminous Blue Variables, e.g. $\eta$ Carinae), or GRB/supernova explosions  \citep{mas03}. 

In massive early-type binaries, the supersonic outflow from one star collides with that of its companion. This collision provokes a strong heating of the shocked wind, leading to the emission of X-rays \citep[see e.g.][]{ste92}. Hints of this high-energy phenomenon were already found two decades ago, when the first X-ray observatories detected overluminosities in hot binaries \citep{pol87, chl91}. However, since additional X-ray emission can have several origins, not restricted to colliding winds (CW), indisputable observational evidence of CW X-ray emission had to await the advent of a new generation of sensitive X-ray facilities (ROSAT, Chandra, \xmm). Indeed, only detailed spectroscopic investigations and careful monitoring in the X-rays can bring to light the actual properties of the hot gas produced by the wind-wind collision. Most notably, phase-locked variations in the X-ray domain are produced by varying separation in eccentric binaries (that changes the intrinsic strength of the collision) or varying line-of-sight opacity as the stars revolve around each other: see e.g. HD\,152248 \citep[O7.5III+O7III, $P$=6d, $e$=0.13;][]{san04}, HD\,93403 \citep[O5.5III+O7V, $P$=15d, $e$=0.23;][]{rau02}, WR\,25 \citep[WN6+O, $P$=208d, $e$=0.5;][]{gam06,cor06}, and $\gamma^2$Vel \citep[WC7+O7.5III, $P$=79d, $e$=0.33;][]{sch04}. Up to now, all studied X-ray colliding wind (XCW) systems belong to our Galaxy - a few XCW candidates have been proposed in 30 Doradus, but solely on the basis of an X-ray overluminosity \citep{por02}, which is insufficient to ascertain the true nature of the emission. Studying the CW phenomenon in other galaxies can provide an important probe of the mass-loss process in different environments with different metallicities. 

\hd, the most peculiar massive star in the Small Magellanic Cloud (SMC), lies on the periphery of the large cluster NGC\,346 associated with the giant H\,{\sc ii} region N\,66. It is a multiple system whose main component, star A, underwent two LBV-like eruptions in 1993-1994, increasing its brightness by up to 3 magnitudes \citep{jon97}. Together with star B, believed to be an early Wolf-Rayet star of the nitrogen sequence, it forms a close eclipsing binary system whose period is 19.3d,  eccentricity is 0.3  and inclination very close to 90$^{\circ}$ \citep{ste97}. Using the ephemeris of the latter authors, star A (resp. B) eclipses its companion at $\phi$=0 (resp. 0.36), whereas periastron and apastron occur at phases 0.09 and 0.59, respectively. A third stellar object, star C (probably an early O-type star), contaminates the light of the system, but it is still unclear if this star is just a line-of-sight coincidence or an object gravitationally bound to the close AB pair. The presence of an additional component (e.g. a close orbiting neutron star) has been proposed, based on a 7-hour periodicity seen in spectral and photometric variations and stochastic polarimetry changes, but is still debated \citep[see e.g. ][]{vil03}. \citet{mof98} infer the presence of a wind-wind collision region in the AB system on the basis of strong emission-line variability. From UV observations, \citet{koe00} and \citet{koe04} conclude that the orientation of the shock cone is such that it wraps around star B, consistent with the notion that star A possesses the more powerful of the two winds. 

The detection of X-ray emission from \hd\ was reported for the first time by \citet{naz02}. Short-term and long-term variations in X-ray brightness were detected \citep{naz02,naz04}. However, these observations were not sufficient to determine the nature of these variations (e.g. variations driven by changes in the shocked ejecta from the 1994 eruption, or phase-dependent changes due to wind-wind collisions in the AB binary), and the origin of the X-ray emission therefore remained uncertain. To resolve this, we acquired additional monitoring of \hd\ with \xmm.

This letter is organized as follows. Section 2 describes the observations while Sections 3 and 4 present the results of the dedicated \xmm\ campaign and their interpretation, respectively. Section 5 gives our conclusions.

\section{Observations and Data Reduction}

During the past few years, \hd\ was observed once with Chandra during 100\,ks and five times\footnote{A sixth observation taken during Rev. 0970 was strongly affected by a flare, rendering the data unusable.} with \xmm\ for approximately 20\,ks each time (see Table \ref{journalobs}). For the latter observations, the three European Photon Imaging Cameras (EPICs) were operated in the standard, full-frame mode (except for the first pn dataset) and a medium filter was used to reject optical light. The observations sampled crucial phases of the 19.3d orbit (Fig. \ref{orbit}); moreover, the last three observations were obtained during the same 19.3d orbit of \hd. To ensure a coherent reduction, we used the Science Analysis System (SAS) software, version~7.0, to re-process all X-ray data. After the pipeline chains, filters recommended by the SAS team were applied \citep[see e.g.][]{naz06}. For consistency, no additional temporal filtering was done.

To each \xmm\ dataset, we applied the SAS source detection algorithm ({\em edetect\_chain}) in a region of 150" radius around \hd, in order to derive the best value of the centroid of the X-ray emission. Since \hd\ appears surrounded by a soft, X-ray bright supernova remnant (SNR, see  \citealt{naz02}), the detection was restricted to the 1.5--10.~keV range, to minimize the contamination from this soft extended source. As a check, we compared the derived value of the count rate with the output of another algorithm (SAS task {\em eregionanalyse}) that simply calculates the number of counts in the region considered\footnote{We used a 25" region centered on \hd, together with a background region of the same area but offset from \hd\ by 22s east in RA (or $-$2040 px in X) and 82" south in DEC (or $-$1640 px in Y). Using an annular background region yields the same results, though noisier.} and corrects them only for the Encircled Energy Fraction (about 0.8 in our case).  Within the error bars, both methods agree with each other so we only present the results from the latter (see Table \ref{journalobs}). 

\section{Results}

\begin{figure}
\plotone{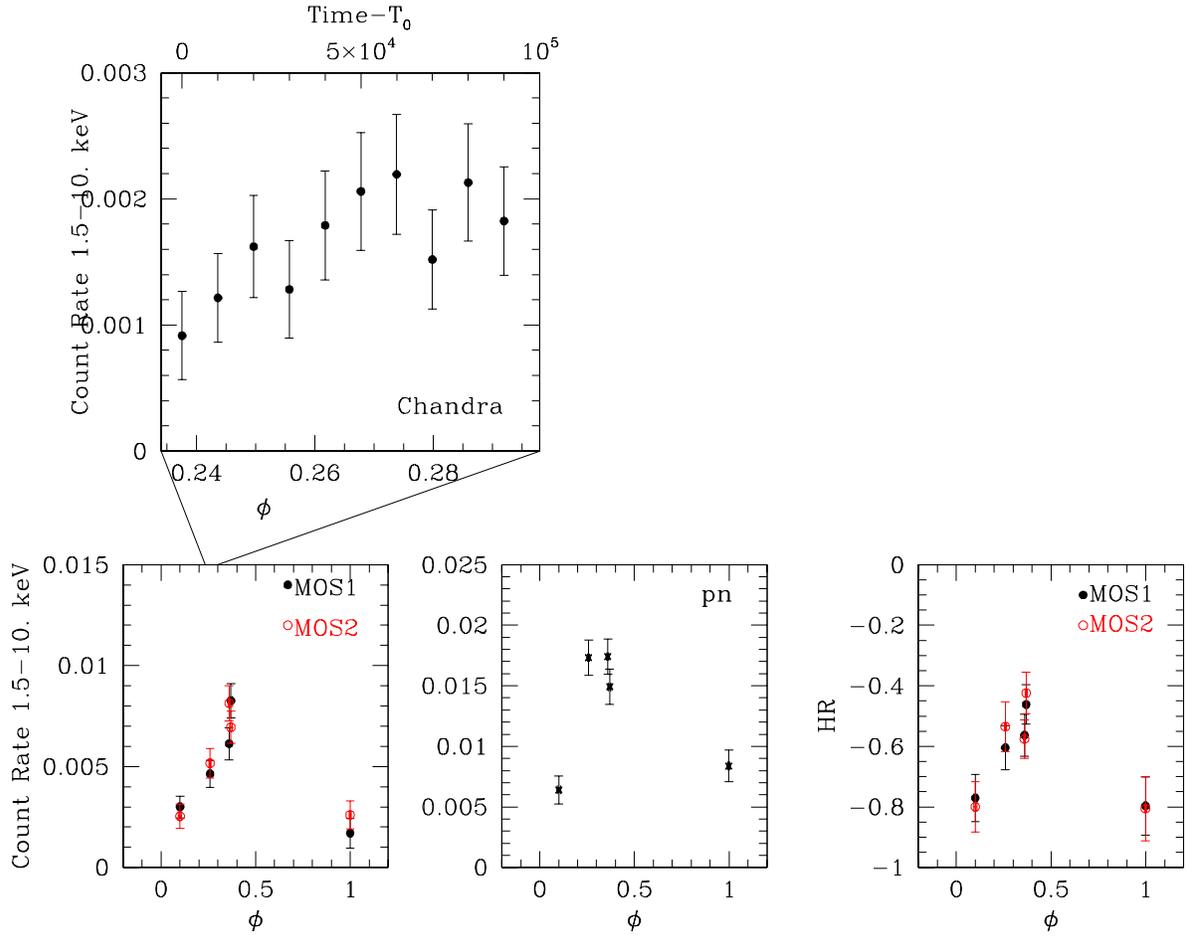}
\caption{Bottom: Evolution of the count rates and hardness ratios with phase from \xmm\ observations (see definitions in Table \ref{journalobs}). Top: Count rate of \hd\ measured during the 100\,ks Chandra observation, in the same energy band. \label{countrate}}
\end{figure}

It must first be noted that two \xmm\ datasets were taken, intentionally, at the same phase ($\phi\sim0.36$). Although these two observations were obtained during \xmm\ Revs. 0157 (2000 Oct.) and 1094 (2005 Nov.), i.e. separated by 5 years or 97 orbital revolutions of the AB system, they present very similar values for the count rate of \hd\ (Fig. \ref{countrate} and Table \ref{journalobs}). This means that post-eruption, epoch-dependent variations are minimal. Indeed, if a significant part of the X-ray emission came from the collision of the fast wind with the slower ejecta associated with the 1994 eruption, a monotonic decrease in the X-ray brightness and a decline in the X-ray temperature with time would be expected as this interaction zone expands and cools. The signature of this slow/fast wind interaction zone was detected as discrete Si\,{\sc{iv}} absorption components at $-$680\,km\,s$^{-1}$ in 1999, only one year before the first X-ray observation, accelerating to $-$710\,km\,s$^{-1}$ in 2000 \citep{koe00,koe01}, and disappearing by 2002. Our \xmm\ observations however indicate that any observable X-ray emission in the 0.5--10 keV band from this interaction is minimal.

\begin{figure}
\plotone{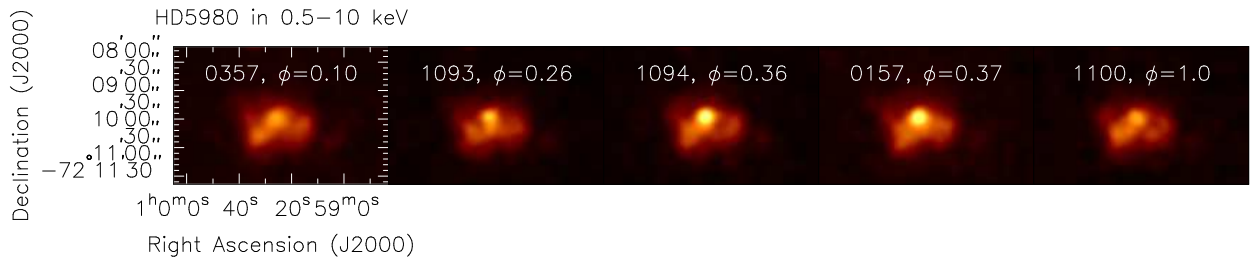}
\caption{Combined MOS images of the area surrounding \hd\ in the 0.5--10.~keV energy range. The data have been binned to get pixels of 2.5" and then smoothed with a gaussian of $\sigma$=3px. \label{allrev}}
\end{figure}

Second, coherent variations with phase are now clearly detected, with X-ray emission steadily increasing towards $\phi=0.36$ (Figs. \ref{countrate} and \ref{allrev}). At that time, the emission also appears slightly harder, though we cannot exclude a constant value of the HR at the 2$\sigma$ level. This bright phase corresponds to the eclipse of star A by its companion, i.e. when the opening of the shock cone, slightly concave with respect to star B, is directed towards the observer (Fig. \ref{orbit}).

Although we cannot directly compare the Chandra and \xmm\ total count rates (because of background contamination, different spatial resolution, and cross-calibration problems), it must be noted that the variations measured by Chandra confirm the trend seen by \xmm. During the 100\,ks of the Chandra observation, the count rate clearly increased from 2 to 4$\times 10^{-3}$\,cts\,s$^{-1}$ in the 0.3--10.~keV band \citep{naz02} or from 1 to 2$\times 10^{-3}$\,cts\,s$^{-1}$ in the 1.5--10.~keV band (Fig. \ref{countrate}). These data covered phases ranging from 0.24 to 0.30. In this interval, the \xmm\ EPIC-MOS lightcurve predicts an increase of $\sim1\times 10^{-3}$\,cts\,s$^{-1}$ in the count rate. Simulations made with PIMMS\footnote{http://heasarc.gsfc.nasa.gov/Tools/w3pimms.html} predict the EPIC MOS and Chandra count rates to be similar in this energy range for a large range of spectral parameters: the variations observed by the two X-ray facilities are thus fully compatible.

Finally, we investigated the X-ray spectrum of \hd. Of course, the faintness of this distant source and the contamination at low energies by the superposed SNR prohibit any detailed study. However, since the spectral properties of the SNR contamination can be determined thanks to the high-resolution Chandra data of \citet{naz02}, a first, general spectral evaluation can be made: our data only reveal a clear, usable excess at high energies for Revs. 0157, 1093 and 1094 (i.e. when \hd\ is the brightest). This high-energy emission was fitted by an absorbed optically-thin plasma model (mekal, in Xspec v11.2.0): compared to the spectral properties reported by \citet{naz02}, only changes of the flux level were detected; within the confidence intervals, the absorption and temperature do not seem to vary in a significant way. However, it must be noted that the spectra are rather noisy and that only large variations would be detected.

\section{Discussion}

The observed variations of the X-ray emission from \hd\ can be qualitatively explained by considering the geometry of the CW region in the close A+B binary.

\begin{figure}
\plotone{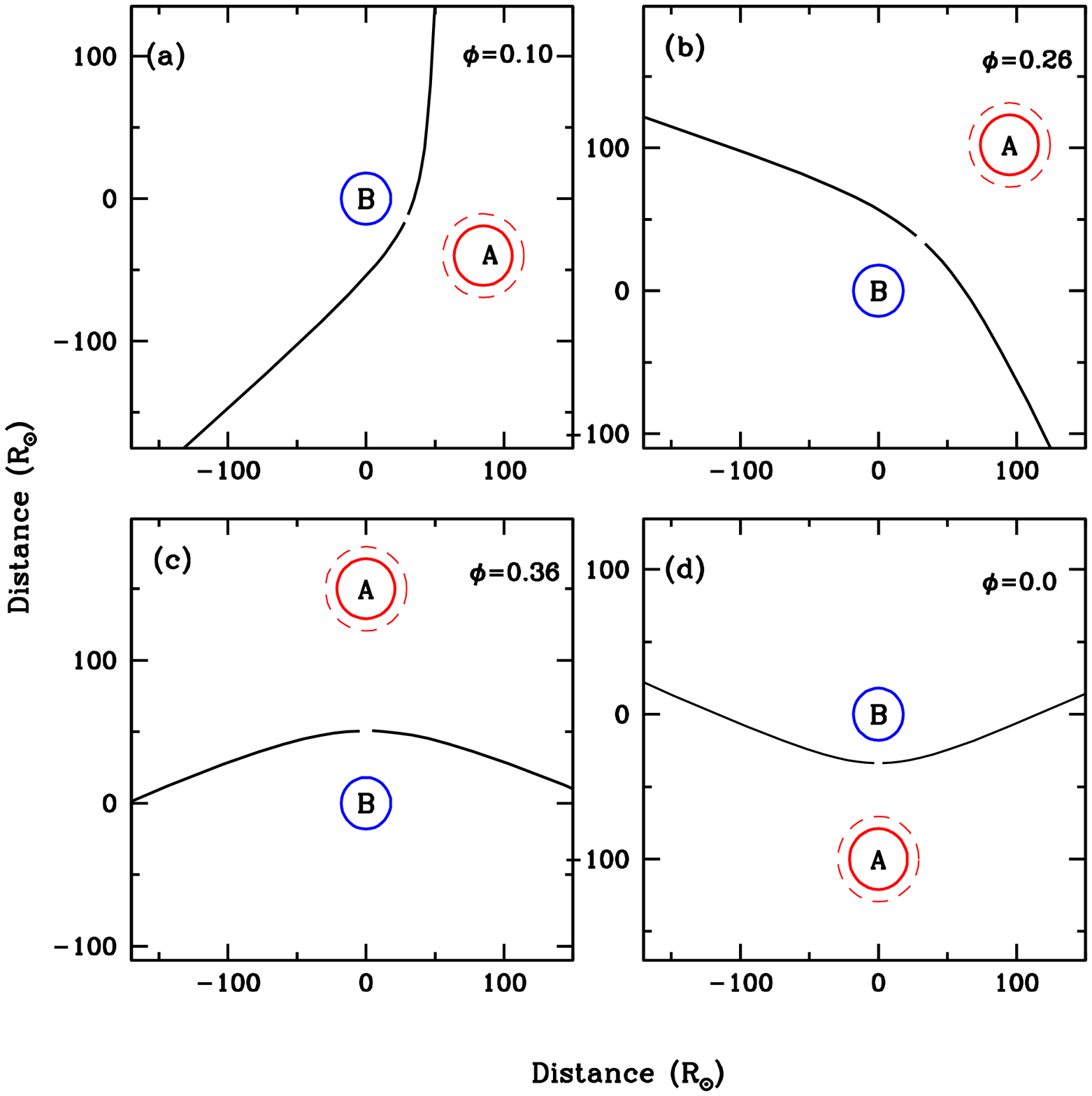}
\caption{Representation of geometry of \hd's wind-wind collision region at the 4 orbital phases for which \xmm\ data are available.  The shape of the shock surfaces was computed under the thin shock approximation \citep{can96} and assuming that $\dot{M}_A$ = 1$\times$10$^{-4}$ M$_\odot$ yr$^{-1}$, $\dot{M}_B$ = 2$\times$10$^{-5}$ M$_\odot$ yr$^{-1}$, v$_\infty$(A)=2000 km s$^{-1}$, and v$_\infty$(B)=2600 km s$^{-1}$.  The broken circle around star A represents the extent of the wind accelerating region \citep{koe06}. \label{orbit}}
\end{figure}

The change in the X-ray flux could be related to absorption. Indeed, at phases close to 0.36, the opening of the shock cone is directed towards the observer and the X-ray emission of the CW region is thus seen through the lower density wind of star B (\citealt{koe04} and Fig. \ref{orbit}). The lower absorption results in an increase of the observed X-ray luminosity. At other phases, the very dense, slower wind of star A strongly absorbs at least part of the CW emission. Such a scenario has been proposed to explain the behaviour of the WR+O binary $\gamma^2$Vel \citep{sch04}. If the X-ray lightcurve of \hd\ is symmetric with a peak at $\phi$=0.36, the width of the lightcurve can be estimated to FWHM=0.26 in phase, which corresponds to a half opening angle of the shock cone $\theta$ of about 46$^{\circ}$ \citep{wil95} and a momentum ratio of 3--4 \citep{ste92}, suggesting that the wind of star A is even stronger than shown in Fig. \ref{orbit}.

However, the available observations do not yet enable us to ascertain that the maximum brightness really occurs at/near $\phi$=0.36. For example, it is conceivable that the count rate actually continues to increase beyond this phase, e.g. to peak at apastron. A similar variation, proportional to the separation between the stars rather than inversely proportional, was observed for Cyg OB2 \#8A \citep[O6I+O5.5III, $P$=22d, $e$=0.24;][]{deb06}. In that case, the variation is explained by a sudden decrease of the wind velocity at periastron because of a strong radiative inhibition/braking or as a result of perturbations associated with tidal bulge interactions. For \hd, such changes of the wind velocity would mean that the X-ray emission should not only be stronger, but also harder towards apastron, as suggested by the available data. 

To distinguish between these two scenarios, additional X-ray observations are necessary in order to complete the lightcurve near apastron. This would enable us to notably determine the position of the brightness peak (apastron or eclipse of A by B ?) and the exact shape of this increase. A detailed spectral analysis is also needed, but it requires better spatial resolution than that of \xmm\ in order to cleanly disentangle \hd\ from the superposed SNR, therefore permitting access to the lower energy range which is more affected by the absorption variations. 

In addition, hydrodynamical modelling of the fast wind/slow wind interaction region and the CW area should be undertaken to quantify their respective strength and expected modulation. In this context, the lack of strong secular variations over 5 years is rather puzzling, since, during this interval, the wind density of star A changed, as witnessed in the UV and optical emission lines \citep{koe04}. However, the X-ray luminosity (in the radiative limit) goes as $L_{X}\propto \dot{M} \times v_{\infty}^{2}$ \citep{ste92}, and since the eruption the terminal velocity of the wind of star A has increased from about 600 km s$^{-1}$ to about 2000 km s$^{-1}$ \citep{koe04}, which offsets the order-of-magnitude decline in the mass loss rate over that interval.  On the other hand, the relative constancy of the X-ray brightness at $\phi=0.36$ despite a significant decline in the mass loss from star A might indicate that star A's wind has mainly changed in the direction perpendicular to the orbital plane than it has in the orbital plane, as might be expected if the mass loss from star A is not spherically symmetric \citep{vil03}. Note however that the observed P-Cygni UV absorption lines indicate that at least some of star A's wind variation occurred near the orbital plane \citep[and references therein]{koe04}.

\section{Summary and Conclusions}

An \xmm\ monitoring campaign of the peculiar massive binary \hd\ has unveiled the variations of its X-ray emission. As two observations taken 5 years apart (a separation of more than 90 binary orbits!) present similar count rates, the main source of X-rays cannot be the fast wind/slow wind interaction following the LBV-like eruption of star A, since its emission is expected to monotonically and rapidly decrease with time. Because individual Wolf-Rayet stars and LBVs are only weak X-ray sources, the high-energy radiation must be associated with the wind-wind collision in the close A+B pair, a fact further supported by the detection of phase-dependent changes. This is the first time that the presence of X-ray emitting gas produced by the collision of winds in a binary has been confirmed outside our Galaxy.

The X-ray emission of \hd\ appears modulated with phase: a clear increase is observed towards the time of the eclipse of star A by its companion. This can be due either to the lower absorption inside the shock cone (a situation reminiscent to that of $\gamma^2$Vel) or to a lower wind velocity at periastron (similar to the behaviour of Cyg OB2 \#8A). Additional data and hydrodynamical modelling are now needed to further distinguish between these two scenarios.

\acknowledgments

YN acknowledges support from the Fonds National de la Recherche Scientifique (Belgium), the PRODEX XMM and Integral contracts, and the visitor's program of the GSFC. GK acknowledges support from DGAPA/PAPIIT IN 119205. AFJM is grateful for financial support from NSERC (Canada) and FQRNT (Quebec).\\

{\it Facilities:} \facility{XMM-Newton (EPIC)}, \facility{CXO (ACIS)}.

\clearpage

\begin{table}
\begin{center}
\caption{Results from the \xmm\ observing campaign. Count rates were evaluated by the simple {\em eregionalyse} task and are given in the 1.5-10.~keV band (in units of $10^{-3}$\,cts\,s$^{-1}$). Hardness Ratios (HR) are defined as $(H-M)/(H+M)$ with $M$ and $H$ the count rates in the 1--2 and 2--10~keV bands, respectively. Phases refer to the ephemeris of \citet{ste97}, while dates are Julian Dates minus 2\,450\,000d.
\label{journalobs}}
{\footnotesize
\begin{tabular}{lccccccc}
\tableline\tableline
Rev. & Date & Duration & $\phi$ & Count rate MOS1 & Count rate MOS2 & Count rate pn & HR (MOS1) \\
\tableline
0157 & 1835.245 & 0.245d & 0.37 & 8.26$\pm$0.85 & 6.95$\pm$0.79 & 14.92$\pm$1.47 & $-0.46\pm0.06$ \\
0357 & 2234.643 & 0.322d & 0.10 & 3.00$\pm$0.54 & 2.54$\pm$0.60 & 6.41$\pm$1.16 & $-0.77\pm0.08$ \\
1093 & 3701.863 & 0.218d & 0.26 & 4.64$\pm$0.67 & 5.16$\pm$0.75 & 17.32$\pm$1.44 & $-0.60\pm0.07$ \\
1094 & 3703.807 & 0.207d & 0.36 & 6.13$\pm$0.80 & 8.14$\pm$0.88 & 17.41$\pm$1.45 & $-0.56\pm0.07$ \\
1100 & 3716.125 & 0.208d & 1.00 & 1.70$\pm$0.73 & 2.60$\pm$0.69 & 8.38$\pm$1.29 & $-0.80\pm0.10$ \\
\tableline
\end{tabular}
}
\end{center}
\end{table}

\clearpage

\clearpage

%\clearpage

\end{document}